\documentclass[aps,pra,reprint,nofootinbib,showpacs,superscriptaddress]{revtex4-1}
\input{Qcircuit}
\usepackage{graphicx} 
\usepackage{hyperref}
\usepackage{amsfonts}
\usepackage{amsmath,amssymb}
\usepackage{mathrsfs}
\usepackage{amsthm}
\usepackage[shortlabels]{enumitem}
\usepackage{dsfont}
\usepackage{bm} 
\usepackage{color}
\usepackage{epstopdf} 
\usepackage{epsfig}

{\rm }

\def\be{\begin{equation}}
 \def\ee{\end{equation}}
 \def\bea{\begin{eqnarray}}
 \def\eea{\end{eqnarray}}
 \def\bes{\begin{eqnarray}}
 \def\ees{\end{eqnarray}}
 \def\bi{\begin{itemize}}
 \def\ei{\end{itemize}} 

\renewcommand{\sec}[1]{\hyperref[sec:#1]{Sec.~\ref{sec:#1}}}

\newcommand{\fig}[1]{\hyperref[fig:#1]{Fig.~\ref{fig:#1}}}

\def\2{\frac{1}{2}}
\def\4{\frac{1}{4}}

\begin{document}

\title{Quantum topological data analysis with continuous variables}

\author{George Siopsis}
\email{siopsis@tennessee.edu}
\affiliation{Department of Physics and Astronomy, The University of Tennessee, Knoxville, Tennessee 37996-1200, U.S.A.}

\date{\today}

\begin{abstract}

I introduce a continuous-variable quantum topological data algorithm. The goal of the quantum algorithm is to calculate the Betti numbers in persistent homology which are the dimensions of the kernel of the combinatorial Laplacian. I accomplish this task with the use of qRAM to create an oracle which organizes sets of data. I then perform a continuous-variable phase estimation on a Dirac operator to get a probability distribution with eigenvalue peaks. The results also leverage an implementation of continuous-variable conditional swap gate. 

\end{abstract}

\maketitle

\section{Introduction}

Extracting useful information from data sets is a difficult task, and in large cases it can be impossible on a classical computer. It is an ongoing field of research to produce quantum algorithms which can analyze data at large scales \cite{bib1, CVQML, PatrickML, 1Qb, Huang2018}. Topological methods for data analysis allow for general useful features of the data to be revealed, and these features do not depend of the representation of the data or any additional noise. This makes topological techniques a powerful analytical tool \cite{Zomorodian2005, Robins1999, Frosini1999, Carlsson2005, Edels2002, Zomorodian2009, Chazal2007, Cohen2007, Basu1999, Basu2003, Basu2008, Niyogi2011, Harker2014, Mischaikow2013}. These methods classically scale with exponential computing time, but have been shown to be a great example of the power of quantum algorithms \cite{bib1, 1Qb, ChuGrover, Huang2018}.

In this work, I follow and build upon the results in \cite{bib1}, focusing on finding the Betti numbers in persistent homology. Persistent homology is a topological method that revolves around representing a space in terms of a simplicial complex and examining the application of a scaled boundary operator. The Betti numbers represent features of the data, such as the number of connected components, holes, and voids. In order to determine the Betti numbers, I use a quantum algorithm which employs the tools of continuous-variable (CV) quantum computation. More specifically, I use quantum principal component analysis (QPCA) to resolve a spectrum of Betti numbers \cite{CVQML}.

A CV quantum system is one that utilizes an infinite-dimensional Hilbert space, where the measurement of variables produces a continuous result. This is a substrate that is being studied extensively, and shown to have applications in generating entanglement, quantum cryptography, quantum teleportation, and quantum computation \cite{CVGaussian, CV2005, CV1999, Menicucci2014, Zhang2006, Menicucci2006, Yokoyama2015, Pysher2011, Takeda2013, Gu2009, Alex2014}. The use of a continuous system provides advantages over a qubit, or discrete-variable (DV) system, such as low cost of optical components, less need for environmental control, and potentially better scaling to larger problems \cite{CVQML, CVGaussian, CV1999, Pysher2011, Takeda2013, Yoshikawa2016, Yokoyama2013, Loock2007, Gu2009}. The CV substrate has also been demonstrated to be more useful in situations with high rate of information transfer such as computing on encrypted data \cite{CVEntangle, Yoshikawa2016, Takeda2013, Yokoyama2013}. These advantages can be very useful when analyzing and extracting information on large volumes of classical data, and as a result make CV quantum computing the natural choice in this setting.

The body of this work starts by discussing persistent homology in a general way, and by setting up the mathematical background of the algorithm, including some useful definitions such as the combinatorial Laplacian and the $k$th Betti number. I then define the use of quantum Random Access Memory (qRAM) which allows a mapping of classical data into a set of quantum states \cite{Lloyd2008, Lloyd20082, Martini2009}. In addition, I outline the process of exponentiation of a Hermitian operator, and arrive at the construction of an oracle that returns the elements of the $k$th Vietoris-Rips complex. Finally, I provide the CV quantum algorithm which uses the process of QPCA, utilizing an implementation of a hybrid \cite{Lloyd2000} exponential conditional swap, to determine the Betti numbers of the system.

The discussion is organized as follows. In Section~\ref{Background}, I discuss in general the steps involved in persistent homology as well as some basics in topological data analysis. In Section~\ref{Setup}, I introduce pertinent mathematical background needed to set up the algorithm. Section~\ref{Tools} describes the usage of qRAM, exponentiation, and the oracle. Section~\ref{QAlgo} outlines the algorithm. I offer a discussion and a conclusion in Section~\ref{Conclusion}.

\section{Background} \label{Background}

   The final goal of topological data analysis, along with the algorithm introduced here, is to determine interesting features of a data set. In this case, the indicator of structure is the Betti numbers, which are a count of topological features. The Betti numbers distinguish between topological spaces based on their connectivity, and are grouped based on dimension. The common notation for Betti numbers is $\beta_{k}$, where for $k=0,1,2$ one has the Betti numbers that correspond to connected components, one-dimensional holes, and two-dimensional voids, respectively.  

\begin{figure}[h]
    \centering
    \includegraphics[width=0.5\textwidth]{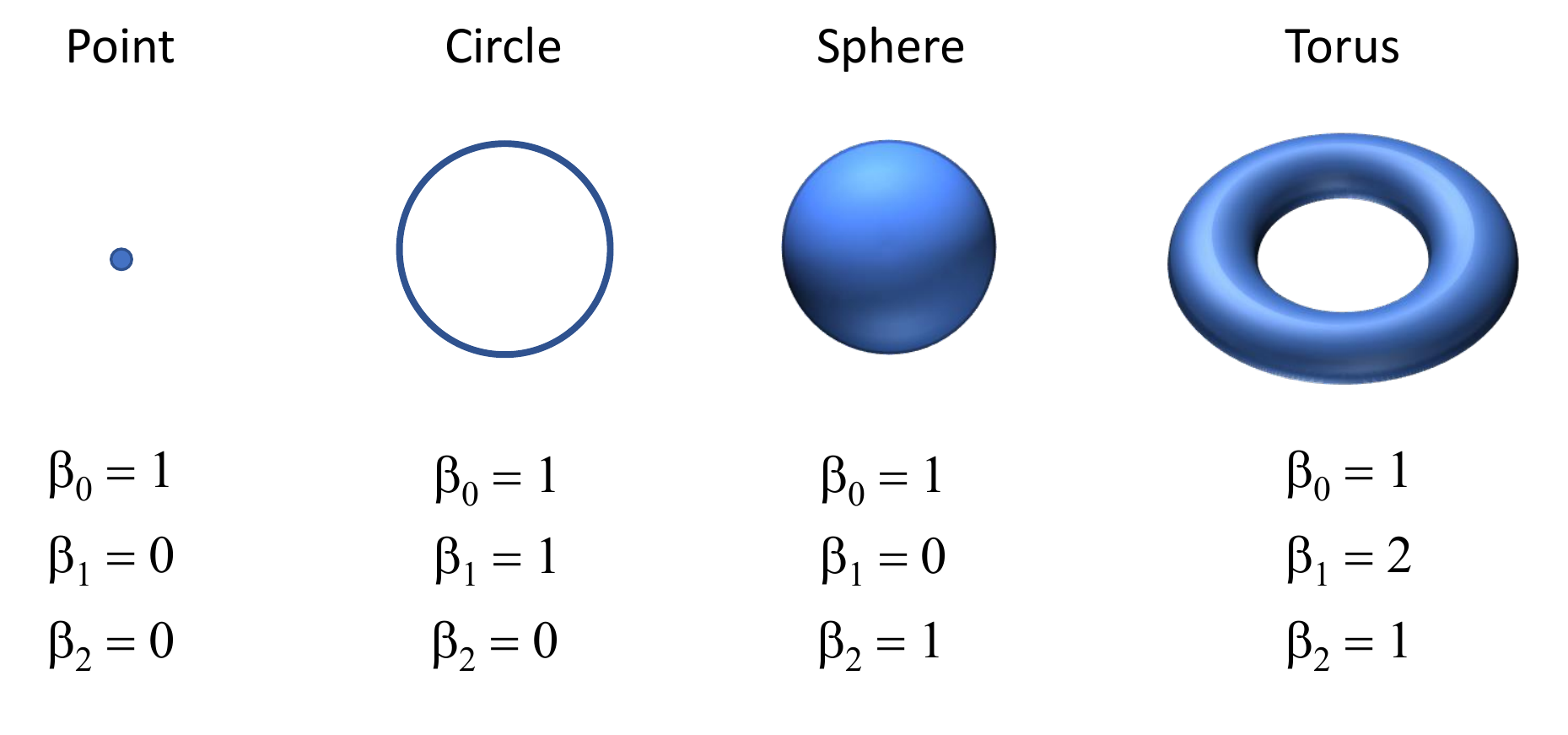}
    \caption{The Betti numbers $\beta_{0,1,2}$ for four example shapes. They are the number of connected components, one-dimensional holes (also called tunnels or handles), and two-dimensional voids, respectively.}
    \label{BettiFig1}
\end{figure}

\begin{figure}[h]
    \centering
    \includegraphics[width=0.45\textwidth]{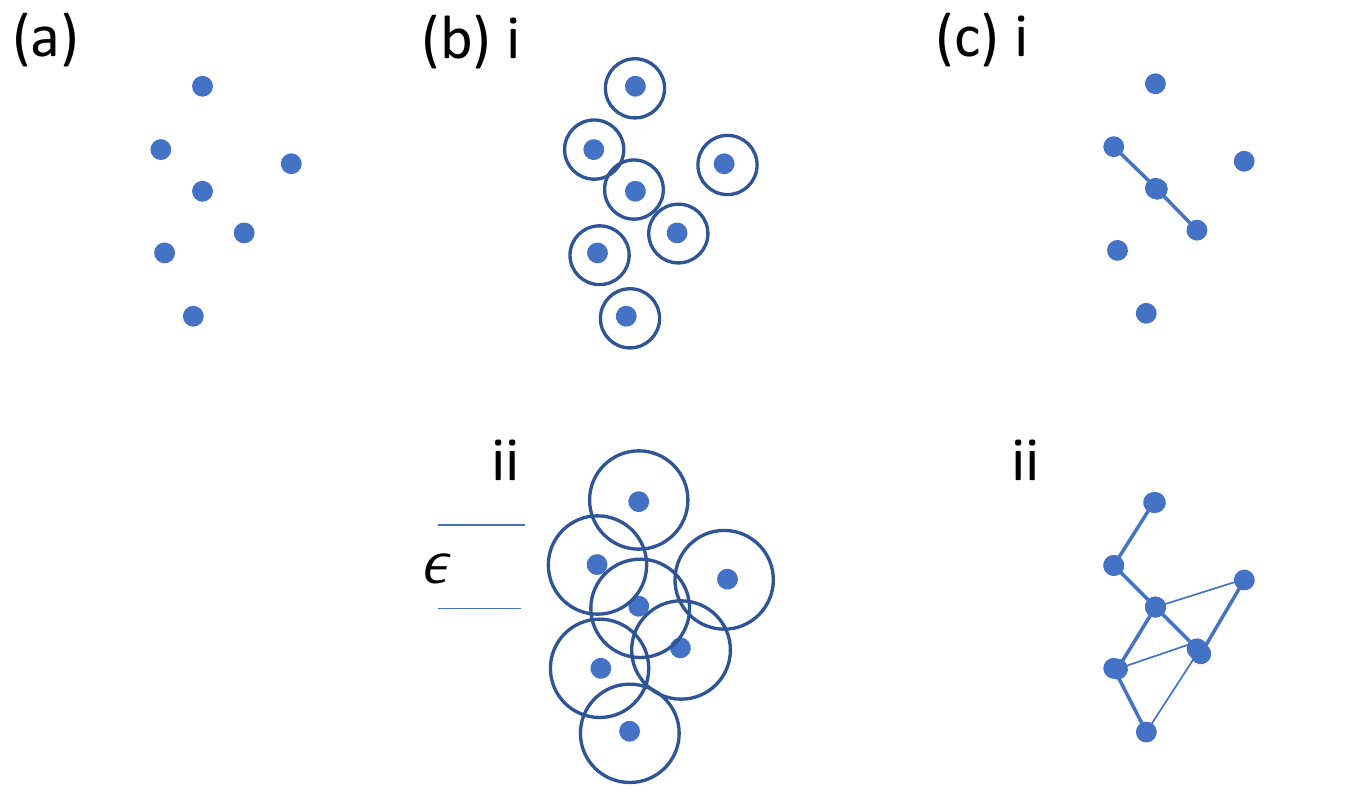}
    \caption{(a) Given data represented by points. (b) For a given distance $\epsilon$, a circle is drawn around each point. (c) Between every two points with contacting circles a line is drawn. These connections are edges of $n$-dimensional shapes (simplices), and the space of simplices in (c) is called a simplicial complex. For two different values of $\epsilon$, as in (b) i, ii, and (c) i, ii, one can get more or less connections between the data points resulting in different topologies. Therefore Betti numbers depend on the initial choice of $\epsilon$. It is useful to vary $\epsilon$ to find interesting structures.}
    \label{BettiFig2}
\end{figure}

   As an example, see Figure~\ref{BettiFig1}, where I consider some simple topological surfaces in one to three dimensions, and list the values of the $k=0,1,2$ Betti numbers. The algorithm introduced in this work allows one to find the Betti numbers after representing some given data in a space of vertices with connecting edges. Starting with the data, one can create this representation in the following steps.
\begin{enumerate}[(a)]
\item Start by allowing each data point to represent a position vector, and place one point at the end of each of these vectors. These points are referred to as the vertices.
\item Next, for a given $\epsilon$ diameter, draw a circle around each vertex in the space.
\item Between every two vertices which have contacting circles (and as such are less than $\epsilon$ distance apart), draw a connecting line.  These connections are edges of $n$-dimensional shapes called simplices, and the space of simplices is called a simplicial complex.
\end{enumerate}
 This process is visualized in Figure~\ref{BettiFig2}. Now, in order to begin to analyze this representation of the data on a quantum computer, they must be first encoded in a quantum state. This is the subject of the next section.

\section{Initialization of algorithm} \label{Setup}

To start, we are given $n$ points in a $d$-dimensional space at position vectors $\bm{v}_i$, $i=1,2,\dots,n$. For simplicity, assume that all points are on the unit sphere, $|\bm{v}_i|=1$. More general sets of points can be considered by extending the discussion in a straightforward, albeit somewhat tedious, manner. For each vector, construct the quantum state
\be \label{Vector} |\bm{v}_i\rangle = \sum_{j=0}^{d-1} (\bm{v}_i)_j |j\rangle, \ee
using $\log_2 d$ qubits. This is analogous to the first part of Figure~\ref{BettiFig2} where data are represented with a series of dots of variable distance from one another.

\begin{figure}[h]
    \centering
    \includegraphics[width=0.5\textwidth]{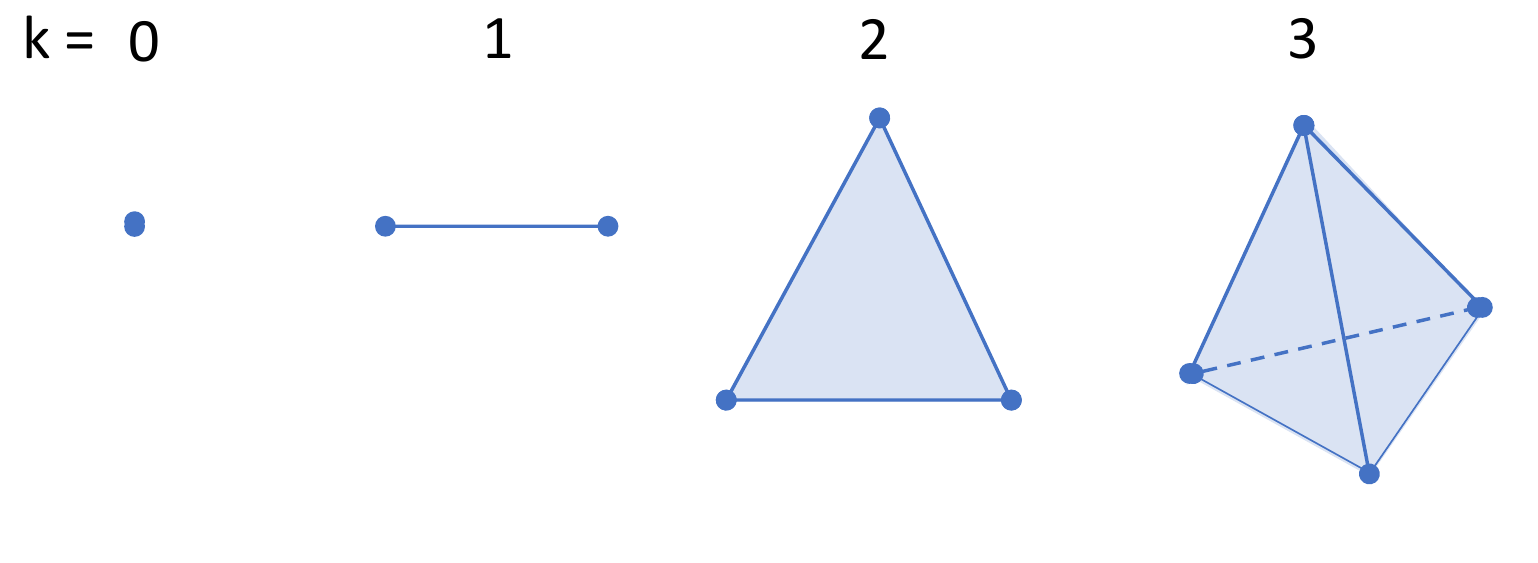}
    \caption{The $k$-simplices for $k=0,1,2,3$. These are a vertex, an edge, a triangle, and a tetrahedron, respectively.}
    \label{BettiFig3}
\end{figure}

\begin{figure}[h]
    \centering
    \includegraphics[width=0.5\textwidth]{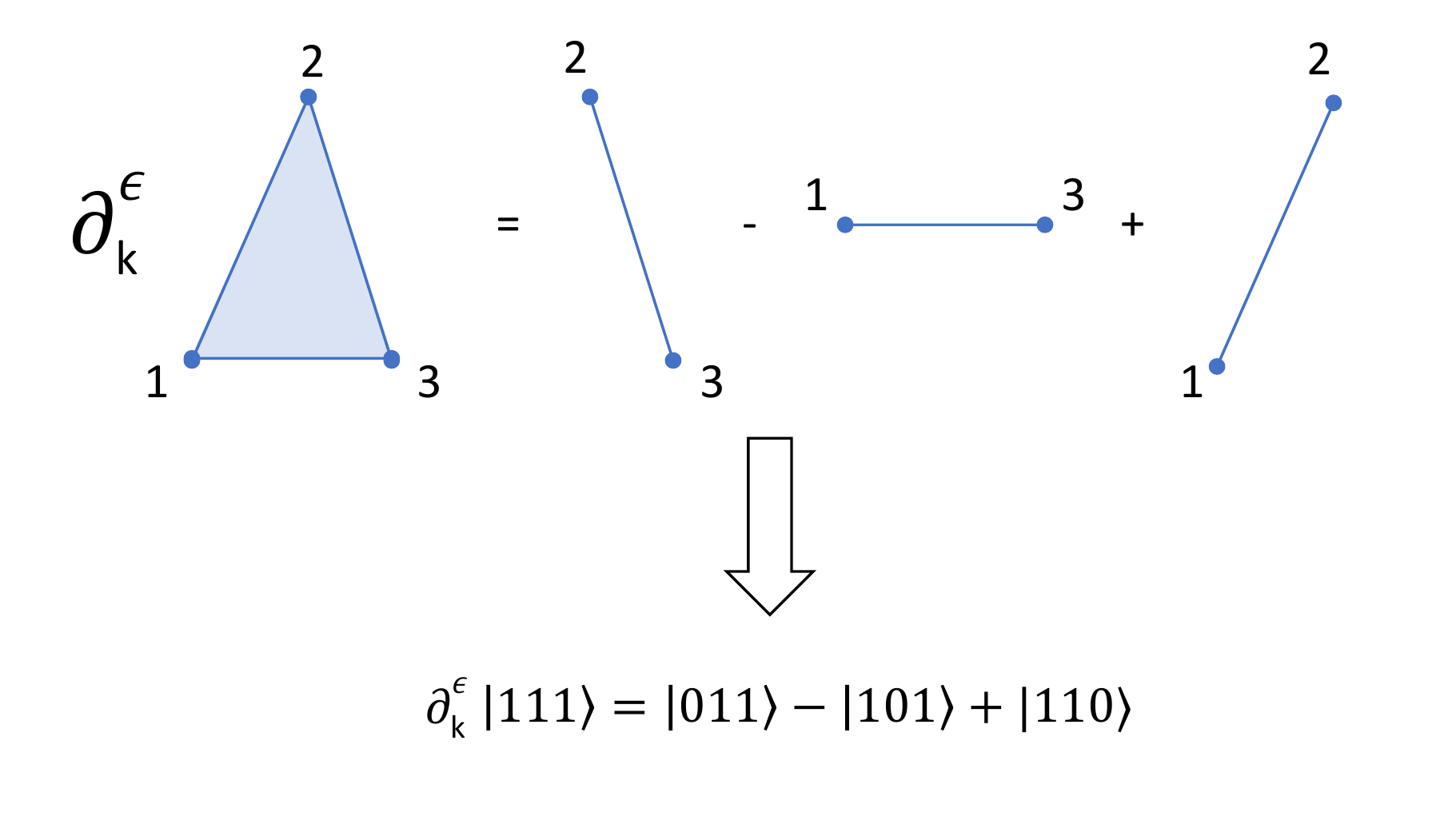}
    \caption{The action of the boundary operator is shown on a $k=2$ simplex. A visual representation of a simplex being broken down into its boundary is depicted above. Its boundary consists of simplices of $k-1=1$. Below is the encoded representation of the boundary operator acting on the 2-simplex. In this encoding a 1 represents a vertex in the corresponding position in the string of bits. The boundary sum is represented by a clockwise rotation around the original simplex, and the negative sign in the result alternates as in Eq.~\eqref{BoundaryOp}.}
    \label{BettiFig4}
\end{figure}

A $k$-simplex $s_k$ is defined as a simplex consisting of $k+1$ vertices at points $\bm{v}_{i_0}, \dots, \bm{v}_{i_{k}}$ connected with $k(k+1)/2$ edges. The first four $k$-simplices are shown in Figure~\ref{BettiFig3}. A simplex can be represented by a string of $n$ bits consisting of $k+1$ $1$s at positions $i_0,\dots, i_k$, and $0$s otherwise (for an example of this representation see Figure~\ref{BettiFig4}). Let $s_k$ denote the number written as this string of $n$ bits in binary notation ($s_k=0,1,\dots, 2^n-1$, if all values of $k$ are considered). The state $|s_k\rangle$ is constructed using $n$ qubits. Thus, simplices are mapped onto basis vectors $|s_k\rangle$.

Next, define the diameter $\mathcal{D} (s_k)$ as the maximum distance between two vertices of the simplex,
\be \label{Diameter} \mathcal{D} (s_k) = \max_{i_l,i_{l'}} |\bm{v}_{i_l} - \bm{v}_{i_{l'}} |. \ee
This allows one to define the Vietoris-Rips complex $S_k^\epsilon$ as the complex consisting of all $k$-simplices with diameter $\mathcal{D} \le \epsilon$, for a given scale $\epsilon$. The construction of the Vietoris-Rips complex is equivalent to the latter pair of steps in Figure~\ref{BettiFig2}, where circles of diameter $\epsilon$ are drawn around each vertex, and then the vertices of contacting circles are connected with edges. The objective of persistent homology is to continuously vary the scale $\epsilon$ until one finds a value which gives the space an interesting structure, as determined by the Betti numbers. The word persistent comes from this varying of $\epsilon$, whereas homology is the algebraic tool that measures the structure of the complex. For the algorithm in the Hilbert space of $n$ qubits, one can define the projection operator
\be P_k^\epsilon =  \sum_{s_k\in S_k^\epsilon} |s_k\rangle\langle s_k|, \ee
onto $S_k^\epsilon$. Evidently, for $\epsilon \ge 2$, all $k$-simplices are included in $S_k^\epsilon$, so one need only consider $\epsilon \in (0,2)$. The parameter $\epsilon$ can be encoded using $m$ qubits as $|x\rangle$, $x=0,1,\dots, 2^m-1$, where $\epsilon = x/2^{m-1}$.

If one removes the $l$th vertex from the $k$-simplex $s_k$, one obtains the $(k-1)$-simplex $s_{k-1}(l)$, $l=0,1,\dots,k$. Evidently,
\be |s_{k-1} (l)\rangle = X_{i_l} |s_k\rangle, \ee
where $X_{i_l}$ is $X$ acting on the $i_l$th qubit. Let us now probe the space to determine whether or not any interesting features are present. This probing is done by the boundary map acting on the space. Define the boundary map $\partial_k$ by
\be \label{BoundaryOp} \partial_k |s_k\rangle = \sum_{l=0}^k (-)^l |s_{k-1} (l)\rangle  = \sum_{l=0}^k (-)^lX_{i_l} |s_k\rangle. \ee
One easily deduces $\partial_k \partial_{k+1} = 0$. To restrict to Vietoris-Rips complexes, introduce
\be \tilde{\partial}_k^\epsilon = P_{k-1}^\epsilon \partial_k P_k^\epsilon. \ee
The action of the boundary operator on a simplex, as well as an example encoding into bits is visualized in Figure~\ref{BettiFig4}.

The entire Hilbert space of $n$ qubits is split into $n+1$ subspaces labeled by $k$. To keep track of this splitting, I introduce a register of $\log_2 n$ qubits to store the state $|k\rangle$ and map
\be |s_k\rangle \mapsto |k\rangle |s_k\rangle, \ee
in parallel. This can be done in $n$ steps as follows. Start with the state $|0\rangle$ for the register. Apply the permutation $\mathcal{P} \ : \ |0\rangle \to |1\rangle \to \dots \to |n-1\rangle \to |0\rangle$ for each digit of $|s_k\rangle$ equal to 1 (using the qubit corresponding to each digit as control). The permutation $\mathcal{P}$ is a 1-sparse matrix and can be implemented efficiently. Thus one applies $\mathcal{P}^k$, so $|0\rangle \to \mathcal{P}^k |0\rangle = |k\rangle$, as desired.

A general state can be written as
\be \label{GenState} |\Psi\rangle = \frac{1}{\sqrt{n}}\sum_{k=0}^{n-1} |k\rangle |\Psi_k\rangle = \frac{1}{\sqrt{n}} \left( \begin{array}{c}
	|\Psi_0\rangle \\ |\Psi_1\rangle \\ \vdots \\ |\Psi_n \rangle
\end{array}\right),\ee
where $|\Psi_k\rangle$ is in the span of $\{ |s_k\rangle \}$.

\begin{figure}[h]
    \centering
    \includegraphics[width=0.5\textwidth]{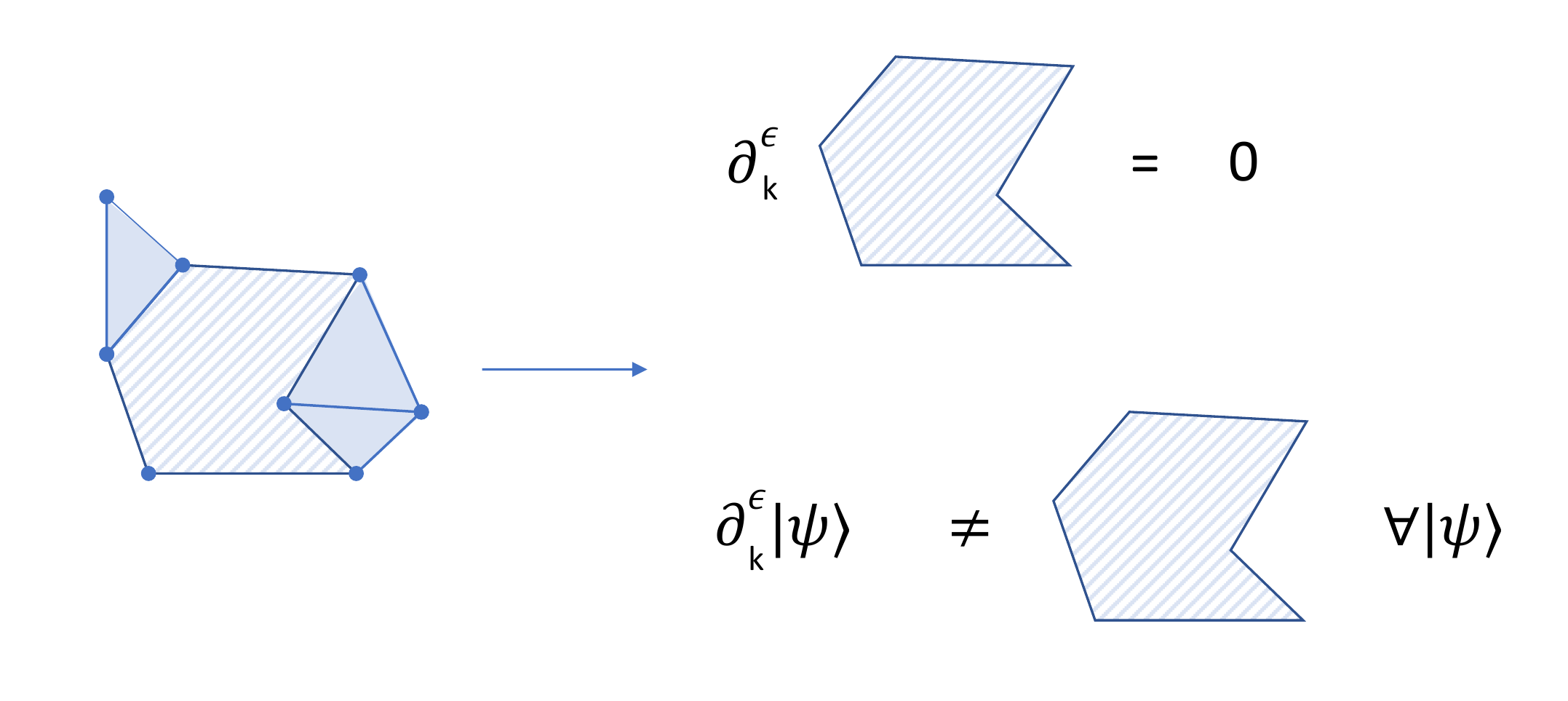}
    \caption{Consider the $k=2$ complex on the left, for a given value of $\epsilon$. In order to show that the striped area is a void, it itself must be boundary-less, and not a boundary for any part of the complex. Fulfillment of these two properties is equivalent to the combinatorial Laplacian \eqref{Laplacian} applied to the stripped area returning zero. Therefore this area would be part of the kernel of the combinatorial Laplacian for $k=2$ contributing to the $\beta_{2}$ Betti number.}
    \label{BettiFig5}
\end{figure}

Define the Dirac operator as the Hermitian matrix
\be\label{eq9} \tilde{B}^\epsilon = \left( \begin{array}{cccccc} 0 & \tilde{\partial}_1^\epsilon & 0 & & & \\ \tilde{\partial}_1^{\epsilon\dagger} & 0 & \tilde{\partial}_2^\epsilon & & &  \\ 0 &\tilde{\partial}_2^{\epsilon\dagger} &0 & & & \\
	& & & \ddots & & \\ & & & &  0 & \tilde{\partial}_n^\epsilon \\
	& & & & \tilde{\partial}_n^{\epsilon\dagger}& 0
\end{array}\right) 
\ee
One easily obtains
\be (\tilde{B}^\epsilon)^2 = \text{diag} \left( \Delta_0^\epsilon , \Delta_1^\epsilon , \dots , \Delta_n^\epsilon \right) \ , \ee
where $\Delta_0^\epsilon = \tilde{\partial}_1^{\epsilon\dagger} \tilde{\partial}_1^\epsilon$, $\Delta_n^\epsilon = \tilde{\partial}_n^{\epsilon\dagger} \tilde{\partial}_n^\epsilon$, and
\be \label{Laplacian} \Delta_k^\epsilon = \tilde{\partial}_k^{\epsilon\dagger} \tilde{\partial}_k^\epsilon + \tilde{\partial}_{k+1}^{\epsilon\dagger} \tilde{\partial}_{k+1}^\epsilon \ , \ k=1,\dots, n-1 \ee
is the combinatorial Laplacian of the $k$th simplicial complex. The output of a $k$th combinatorial Laplacian being zero tells us exactly that we have found a space which is boundary less and also not a boundary itself. The number of these features in our data is the Betti number. Therefore, one can say that the dimension of the kernel of $\Delta_k^\epsilon$ is the $k$th Betti number,
\be \beta_k = |\ker \Delta_k^\epsilon |. \ee
An example of determining a void in a $k=2$ complex is shown in Figure~\ref{BettiFig5}. The total number of voids is the Betti number $\beta_{2}$.

To summarize, when starting with a data set, the following steps are needed in order to perform persistent homology using the quantum algorithm presented in this paper.
\begin{enumerate}[(a)]
\item Start with $n$ points called vertices in a $d$-dimensional space defined by a set of position vectors. For each vector, the quantum state \eqref{Vector} is constructed.
\item Using the diameter $\mathcal{D} (s_k)$ defined in \eqref{Diameter}, the vertices are connected to form the Vietoris-Rips complex $S_k^\epsilon$ consisting of all $k$-simplices of diameter $\leq \epsilon$.
\item The space is then split into subspaces labeled by $k$, consisting of all $k$-simplices. A general state is then constructed which spans all of these subspaces, and is given in Eq.~\eqref{GenState}.
\item In order to probe the space for interesting structures, use the boundary map \eqref{BoundaryOp}. To determine if a region of the space is one of the features which are tallied up to become the Betti numbers (Holes, Voids, etc.), this region must be boundary-less and also not a boundary of any other part of the space, as shown in Fig.~\ref{BettiFig5}. These two properties are satisfied when the action of the combinatorial Laplacian \eqref{Laplacian} returns zero. The combinatorial Laplacians of $k$th order are the elements of the diagonal matrix which is the square of the Dirac operator \eqref{eq9}.
\item In order to apply this operator in the CV algorithm presented here, and to construct some of the states mentioned above, some additional mathematical tools are required, and outlined in the following section.
\end{enumerate}

\section{Mathematical Tools} \label{Tools}

In this section, the tools needed in the CV quantum algorithm are outlined, and the way they are used and implemented is discussed.

In order to complete the quantum algorithm, We assume that we are equipped with a qRAM \cite{Lloyd2008, Lloyd20082, Martini2009} which, given an input state $|i\rangle |\bm{0}\rangle$, produces the output state $|i\rangle |\bm{v}_i\rangle$ in quantum parallel,
\be \text{QRAM} \ : \ \sum_{i=0}^{n-1} a_i |i\rangle |\bm{0}\rangle \longrightarrow \sum_{i=0}^{n-1} a_i |i\rangle |\bm{v}_i\rangle. \ee
%
%
Another useful tool which will be used in the algorithm is the exponentiation of an operator. Given a Hermitian operator $\bm{A}$, and a resource qumode of quadratures $(q_{\mathcal{R}}, p_{\mathcal{R}})$, it is necessary to apply
\be\label{eq14} e^{ip_{\mathcal{R}}\bm{A} t}~, \ee
in parallel. To this end, I will use the exponential swap operator
\be\label{eqexpS} e^{i\theta p_{\mathcal{R}}\mathcal{S}}~, \ee
where $\mathcal{S}$ is the swap operator. Its implementation is discussed in appendix \ref{sec:A} (and differs from the one given in \cite{CVQML}). While the body of this work uses a CV quantum algorithm, the method of implementing the exponential conditional swap also uses single photon qubits in a hybrid approach \cite{Lloyd2000}. The latter can also be implemented using CV systems in the dual rail representation.

Then we form the state $\rho_{\bm{A}} = \frac{\bm{A}}{\text{tr} \bm{A}}$ (assuming $\text{tr} \bm{A} \ne 0$), and apply \eqref{eqexpS} on the combined system of $|\Psi\rangle$ and $\rho_{\bm{A}}$, for a short time $\delta t$. After tracing out the auxiliary mode $\rho_{\bm{A}}$, we obtain
\bea &&\text{tr}_{\rho_{\bm{A}}} \left( e^{i\delta t p_{\mathcal{R}}\mathcal{S}} |\Psi\rangle \langle\Psi| \otimes \rho_{\bm{A}} e^{-i\delta t p_{\mathcal{R}}\mathcal{S}} \right) \nonumber\\ && = e^{i\delta t p_{\mathcal{R}} \rho_{\bm{A}}} |\Psi\rangle \langle\Psi| e^{-i\delta t p_{\mathcal{R}} \rho_{\bm{A}}} + \mathcal{O} \left( (\delta t)^2 \right) \nonumber\\
 && =e^{i\delta t (\text{tr} \bm{A}) p_{\mathcal{R}} \bm{A}} |\Psi\rangle \langle\Psi| e^{-i\delta t (\text{tr} \bm{A}) p_{\mathcal{R}} \bm{A}} + \mathcal{O} \left( (\delta t)^2 \right). \eea
 By repeating this $\frac{t}{\delta t (\text{tr} \bm{A})}$ times, an approximation to the desired operator \eqref{eq14} is  obtained.

We also assume we are in possession of a quantum oracle $\mathcal{O}_k^\epsilon$ that acts on $|s_k\rangle |\psi\rangle$ in parallel, flipping the last qubit if $s_k \in S_k^\epsilon$, otherwise doing nothing,
\be \mathcal{O}_k^\epsilon |s_k\rangle |\psi\rangle = \left\{ \begin{array}{ccc}
|s_k\rangle \otimes X|\psi\rangle & , & s_k\in S_k^\epsilon \\
|s_k\rangle |\psi\rangle & , & s_k \notin S_k^\epsilon 
\end{array}\right. \ee
This oracle can be implemented in $\mathcal{O} (k^2)$ steps. If we choose $|\psi\rangle = |-\rangle$, where $X|\pm\rangle = \pm|\pm\rangle$, then the last qubit decouples and the oracle is a unitary acting on $|s_k\rangle$ as
\be \mathcal{O}_k^\epsilon |s_k\rangle = \left\{ \begin{array}{ccc}
	-|s_k\rangle & , & s_k\in S_k^\epsilon \\
	|s_k\rangle & , & s_k \notin S_k^\epsilon 
\end{array}\right. \ee
To construct the oracle, first we construct the state
\be \frac{1}{n}\sum_{i,j=0}^{n-1} |i\rangle |j\rangle, \ee
by making two copies of the state $\frac{1}{\sqrt{n}} \sum_i |i\rangle$. To this state we attach $|\bm{0}\rangle \in \mathbb{C}^d$ as well as a qubit in the state $|+\rangle$. We then query qRAM to obtain
\be  \frac{1}{n}\sum_{i,j=0}^{n-1} |i\rangle |j\rangle |\bm{v}_i\rangle |+\rangle. \ee
Then we use the last qubit as control to apply the swap operator and obtain
\be  \frac{1}{\sqrt{2} n}\sum_{i,j=0}^{n-1} |i\rangle |j\rangle \left( |\bm{v}_i\rangle |0\rangle + |\bm{v}_j\rangle |1\rangle \right). \ee
Then we measure $X$ on the last qubit. If the outcome is $-1$, the state collapses to (unnormalized)
\be  \sum_{i,j=0}^{n-1} |i\rangle |j\rangle \left( |\bm{v}_i\rangle  - |\bm{v}_j\rangle \right) |-\rangle. \ee
We then add ancillae and copy the labels $i$ and $j$ on them, respectively. We obtain the state (ignoring the last qubit which has decoupled)
\be  \sum_{i,j=0}^{n-1} |i\rangle |j\rangle |i\rangle_A |j\rangle_A \left( |\bm{v}_i\rangle  - |\bm{v}_j\rangle \right). \ee
Tracing out the ancillae and the last qubit, we obtain the (unnormalized) state
\be H = \sum_{i,j=0}^{n-1} |\bm{v}_i - \bm{v}_j |^2 |i\rangle\langle i| \otimes |j\rangle\langle j|. \ee
It is a Hermitian operator that can be implemented as $e^{itH}$, as discussed above. Its eigenvalues are the distances between points. Also any function of $H$ can be implemented; in particular, the step function $\theta (\epsilon^2 - H)$, that tests whether $s_k \in S_k^\epsilon$; hence the oracle.

\section{Quantum topological data analysis for CVs} \label{QAlgo}

The algorithm requires a register of $m$ qubits to record $\epsilon$ as $|x\rangle$ with $\epsilon = x/2^{m-1}\in (0,2)$. Suppose $\epsilon$ is fixed (a condition that can be relaxed to include a filtration). Let us start with the initial state that includes all $k$-simplices equally weighted,
\be\label{eq18} |\Psi^{(0)}\rangle = \frac{1}{\sqrt{n}}\sum_{k=0}^{n-1} |k\rangle |\Psi_k^{(0)}\rangle = \frac{1}{\sqrt{n}} \left( \begin{array}{c}
	|\Psi_0^{(0)}\rangle \\ |\Psi_1^{(0)}\rangle \\ \vdots \\ |\Psi_n^{(0)} \rangle
\end{array}\right),\ee
where
\be |\Psi_k^{(0)}\rangle = \frac{1}{\sqrt{( {n\atop k+1})}}\sum_{s_k} |s_k\rangle. \ee
The initial state can be constructed from the state $|0\rangle |s\rangle$, where the register $|0\rangle$ consists of $\log_2 n$ qubits, and
\be |s\rangle = \frac{1}{2^{n/2}} \sum_{y=0}^{2^n-1} |y\rangle, \ee
consists of $n$ qubits. By using each qubit in the state $|s\rangle$ as control to apply the permutation $\mathcal{P}$ on the register, we arrive at the desired initial state \eqref{eq18}.

From the initial state \eqref{eq18}, we construct an approximation to the state
\be |\Psi^\epsilon\rangle = \frac{1}{\sqrt{n}}\sum_{k=0}^{n-1} |k\rangle |\Psi_k^\epsilon\rangle = \frac{1}{\sqrt{n}} \left( \begin{array}{c}
	|\Psi_0^\epsilon\rangle \\ |\Psi_1^\epsilon\rangle \\ \vdots \\ |\Psi_n^\epsilon \rangle
\end{array}\right),\ee
where
\be |\Psi_k^\epsilon\rangle = \frac{1}{\sqrt{|S_k^\epsilon|}} \sum_{s_k^\epsilon\in S_k^\epsilon} |s_k^\epsilon\rangle, \ee
using Grover's search algorithm \cite{Grover}, with the aid of the oracle.

Notice that the action of the Dirac operator \eqref{eq9} simplifies, because all projection operators $P_k^\epsilon$ act as the identity on $|\Psi_k^\epsilon\rangle$. Therefore, one could instead consider the simpler operator
\be\label{eq9a} B = \left( \begin{array}{cccccc} 0 & \partial_1 & 0 & & & \\ \partial_1^{\dagger} & 0 & \partial_2 & & &  \\ 0 &\partial_2^{\dagger} &0 & & & \\
	& & & \ddots & & \\ & & & &  0 & \partial_n \\
	& & & & \partial_n^{\dagger}& 0
\end{array}\right) 
\ee
My goal is to compute the eigenvalues of $B$. For Betti numbers, I am interested in the frequency of occurrence of the zero eigenvalue which yields the dimension of the kernel of the combinatorial Laplacian. I will compute the eigenvalues using QPCA, as discussed in \cite{CVQML}, which is a more specific implementation than the original work in \cite{bib1} which cites the use of general Hamiltonian simulation.

Let us attach a squeezed resource qumode in the state (unnormalized)
\be \int dp_{\mathcal{R}} e^{-p_{\mathcal{R}}^2/(2s)} |p_{\mathcal{R}}\rangle \ee
and apply the unitary
\be\label{Unitary} e^{i\gamma p_{\mathcal{R}} B} \ee
where $\gamma$ is a parameter that can be adjusted at will. This unitary is of the form \eqref{eq14}, except that $\text{tr} B = 0$. We need to regulate $B$, by adding $\alpha \mathbb{I}$, where $\alpha$ is arbitrary. The eigenvalues are shifted by $\alpha$, and $\text{tr} (B+\alpha \mathbb{I}) \ne 0$.

Suppose that the eigenvalue problem of $B+\alpha \mathbb{I}$ is
\be (B + \alpha \mathbb{I})|e_i\rangle = \lambda_i |e_i\rangle \ee
and the state is expanded as
\be |\Psi^\epsilon\rangle = \sum_i a_i |e_i\rangle \ee
Then we obtain
\be \sum_i a_i \int dp_{\mathcal{R}} e^{i\gamma p_{\mathcal{R}} \lambda_i} e^{-p_{\mathcal{R}}^2/(2s)} |e_i\rangle |p_{\mathcal{R}}\rangle \ee
A measurement of the quadrature $q_{\mathcal{R}}$ of the resource qumode with homodyne detection projects the state onto
\be \sum_i a_i e^{-s (\gamma \lambda_i - q_{\mathcal{R}})^2 /2} |e_i\rangle |q_{\mathcal{R}}\rangle \ee
with the probability distribution
\be P(q_{\mathcal{R}}) \propto \sum_i |a_i|^2 e^{-s (\gamma \lambda_i - q_{\mathcal{R}})^2} \ee
consisting of peaks at the eigenvalues. If one is interested in distinguishing between eigenvalues, one ought to choose sufficiently large parameters $s$ and $\gamma$ so that the width of each peak, $1/(\gamma\sqrt{s})$, is narrow enough. From this probability distribution, one can deduce all Betti numbers.

\section{Discussion and Conclusion} \label{Conclusion}

In this work, I discussed a quantum algorithm for topological data analysis using the method of persistent homology. The algorithm was designed using qRAM as well as a continuous-variable substrate to take advantage of a continuous output from which Betti numbers can be calculated. I also examined the use of a continuous-variable exponential conditional swap operation which is outlined in more detail in the Appendix. As in the discrete-variable case \cite{bib1}, although the matrix \eqref{eq9a} is exponentially large ($O(2^n)$) the size of the required qRAM is small. This provides an advantage over other algorithms that require a large qRAM \cite{Harker2014, Ghrist2008, Kozlov2008}.

In general, the use of discrete-variable quantum algorithms for topological data analysis is something that has been used before \cite{bib1, 1Qb, ChuGrover, Huang2018}. The work presented here provides new tools for continuous-variable systems as well as a direct circuit implementation of one of those tools.

In the algorithm presented here, I used a subset of phase estimation called principal component analysis in order to determine the eigenvalues of the exponential operator. This method is a natural fit for the continuous-variable framework discussed here, but there are other methods which have been examined. For example, a hybrid approach which uses a qumode as well as a mixed state of qubits \cite{NinaPhase}. There is also the well-understood purely-qubit phase estimation \cite{Nielson2000}, but this approach would require many copies of the unitary \eqref{Unitary}, greatly increasing any resource costs as a result. Note that the algorithm used in this work is largely an adaptation of the exponentiation and phase estimation of ref.\ \cite{CVQML}. The discussion of resource costs in that work can then be sufficiently translated to the present algorithm.


\acknowledgments
   I would like to thank T.\ Kalajdzievski, S.\ Lloyd, P.\ Rebentrost, and C.\ Weedbrook for interesting discussions and helpful suggestions.

\bibliographystyle{apsrev}
\bibliography{TDD}

\appendix
\section{Exponential conditional swap}
\label{sec:A}

Here I discuss the implementation of the exponential swap operator \eqref{eqexpS} conditioned on the quadrature $p_\mathcal{R}$ of a resource mode. Let us concentrate on qubits labeled as $i$ and $j$ that we wish to swap, $S_{ij} |x\rangle_i |y\rangle_j = |y\rangle_i |x\rangle_j$. Each qubit consists of a pair of qumodes in single-photon states, $|10\rangle = a^\dagger |00\rangle$, $|01\rangle = b^\dagger |00\rangle$, which are identified with the computational basis vectors $|0\rangle$, $|1\rangle$, respectively.

Let us introduce the controlled rotations
\be \text{CR}_{ij}^{X} (\theta) = \frac{\mathbb{I}_i + Z_i}{2} +  \frac{\mathbb{I}_i - Z_i}{2} e^{i\theta X_j} = e^{i\theta b_i^\dagger b_i (a_j^\dagger b_j + a_j b_j^\dagger)}\ee
where $\text{CR}_{ij}^{X} (\frac{\pi}{2})$ is the CNOT gate with control (target) the $i$th ($j$th) qubit,
\be \text{CR}_{ij}^{Y} (\theta) = \frac{\mathbb{I}_i + Z_i}{2} +  \frac{\mathbb{I}_i - Z_i}{2} e^{i\theta Y_j} = e^{\theta b_i^\dagger b_i (a_j^\dagger b_j - a_j b_j^\dagger)}\ee
and
\be \text{CR}_{ij}^{Z} (\theta) = \frac{\mathbb{I}_i + Z_i}{2} +  \frac{\mathbb{I}_i - Z_i}{2} e^{i\theta Z_j} = e^{i \theta b_i^\dagger b_i (a_j^\dagger a_j - b_j^\dagger b_j)}\ee
where $\text{CR}_{ij}^{Z} (\frac{\pi}{2})$ is the CZ gate. They can be constructed using quartic phase gates \cite{CVQML}.

The swap gate can be transformed into a CZ gate using
\be \text{SWAP}_{ij} = U_{ij} \cdot \text{CZ} \cdot U_{ij}^\dagger \ , \ U_{ij} = \text{CNOT}_{ji}\cdot \text{CR}_{ij}^Y (\frac{\pi}{4}) \ee
The CZ gate can be conveniently implemented by introducing an ancillary qubit in the state $|0\rangle_A \equiv a_A^\dagger |00\rangle_A$, and using
\be \text{CZ}_{ij} |0\rangle_A = \text{CCNOT}_{ij,A} \cdot Z_A\cdot \text{CCNOT}_{ij,A} |0\rangle_A \ee
We deduce
\bea e^{it p_{\mathcal{R}} \text{SWAP}_{ij}} |0\rangle_A &=& U_{ij}\cdot \text{CCNOT}_{ij,A} \cdot e^{it p_{\mathcal{R}}\cdot  (a_A^\dagger a_A - b_A^\dagger b_A)} \nonumber\\
&& \cdot\text{CCNOT}_{ij,A}\cdot U_{ij}^\dagger |0\rangle_A\eea
If the system contains multiple qubits, then the total swap operator $\mathcal{S}$ is a product of swap operators for individual qubits,
\be \mathcal{S} = \prod_i \text{SWAP}_{ii'} \ee
The above result can be straightforwardly extended. We obtain
\be\label{eqexpS2} e^{it p_{\mathcal{R}} \mathcal{S}} |0\rangle_A = \mathcal{U}  \cdot e^{it p_{\mathcal{R}}\cdot  (a_A^\dagger a_A - b_A^\dagger b_A)}
 \cdot \mathcal{U}^\dagger |0\rangle_A\ee
 where
 \be \mathcal{U} = \prod_i U_{ii'}\cdot \text{CCNOT}_{ii',A} \ee
The circuit for the exponential conditional swap operator \eqref{eqexpS2} for a system of two qubits is shown below.
\[ \ \ \ \ \ \ \ \ \ 
\Qcircuit @C=1em @R=.7em {
	\lstick{1}  &\targ   &\ctrl{1}   &\ctrl{1}  &\qw &\qw &\qw &\ctrl{1} &\ctrl{1} &\targ &\qw\\
	\lstick{1'}  &\ctrl{-1}   &\gate{e^{i\frac{\pi}{4}} Y}  &\ctrl{4} &\qw   &\qw &\qw &\ctrl{4} &\gate{e^{-i\frac{\pi}{4}} Y} &\ctrl{-1} &\qw\\
	\lstick{2}  &\targ   &\ctrl{1}   &\qw &\ctrl{1}  &\qw &\ctrl{1} &\qw &\ctrl{1} &\targ &\qw\\
\lstick{2'}  &\ctrl{-1}   &\gate{e^{i\frac{\pi}{4}} Y}  &\qw &\ctrl{2} &\qw   &\ctrl{2} &\qw &\gate{e^{-i\frac{\pi}{4}} Y} &\ctrl{-1} &\qw\\
	\lstick{\ket{p_{\mathcal{R}}}} & \qw & \qw &\qw &\qw &\ctrl{1} &\qw &\qw &\qw &\qw &\qw \\
	\lstick{\ket{0}_A} & \qw & \qw & \targ &\targ & \gate{e^{itp_{\mathcal{R}} Z} }&\targ &\targ  &\qw &\qw &\qw\\
}
\]

\end{document}